\documentstyle[equation]{article}

\def\p{\partial}
\def\dfrac#1#2{{\displaystyle\frac{#1}{#2}}}
\def\lb#1{\label{#1} }
\def\non{\nonumber}
\def\I{{\cal I}}
\def\J{{\cal J}}
\def\L{{\cal L}}
\def\be{\begin{equation}}
\def\ee{\end{equation}}
\def\ba#1{\begin{array}{#1}}
\def\ea{\end{array}}
\def\bea{\begin{eqnarray}\samepage}
\def\eea{\end{eqnarray}}
\def\bse{\begin{subequations}}
\def\ben{\begin{eqalignno}}
\def\een{\end{eqalignno}}
\def\ese{\end{subequations}}
\def\i{{\rm i}}
\def\ki{\hat{\imath}}
\def\hgamma{\hat{\gamma}}
\def\hrho{\hat{\rho}}
\def\hF{\hat{F}}
\def\hp{\hat{\p}}
\def\hE{\hat{E}}
\def\hB{\hat{B}}

\def\hLambda{\hat\Lambda}
\def\sqr{{\scriptstyle\sqrt{|g|}}}

\begin{document}
\title{\bf Born-Infeld electrodynamics:
 Clifford number and spinor representations}
\author{Alexander A. Chernitskii\\
{\small Saint Petersburg Electrotechnical University}\\
{\small Prof. Popov Str. 5, 197376 Russia}\\ {\small aa@cher.etu.spb.ru}}
\date{}
\maketitle

\begin{abstract}
Clifford number formalism for Maxwell equations is considered. The
Clifford imaginary unit for space-time is introduced as coordinate
independent form of fully antisymmetric fourth-rank tensor. The
representation of Maxwell equations in massless Dirac equation form
is considered; we also consider two approaches to the invariance
of Dirac equation in respect of the Lorentz transformations. According to the
first approach, the unknown column is invariant and according to the second
approach it has the transformation properties known as spinorial ones.
Clifford number representation for nonlinear electrodynamics equations
is obtained. From this representation, we obtain the nonlinear like
Dirac equation which is the form of nonlinear electrodynamics equations. As
a special case we have the appropriate representations for
 Born-Infeld nonlinear electrodynamics.
\\[1.5ex]
2000 Mathematics Subject Classification:
35Q60, 15A66.
\end{abstract}

\section{Introduction}
At present Born-Infeld nonlinear electrodynamics arouse considerable interest
for various points of view. There are some propositions which allow
considering this nonlinear electrodynamics as an unique field model.
This model, considered by Born and Infeld \cite{BornInfeld},
is obtained from some variational principle proposed by
Eddington \cite{Eddington1924}. This model appear also in
the string theory \cite{FradkinAndTseytlin1985}.
It was shown in \cite{ChernikovAndShavokhina1986} that
the electrodynamical part of the Einstein's unified field model,
with non-symmetrical metric \cite{EinsteinAndCaufmann}, is equivalent to
Born-Infeld electrodynamics.
The characteristic equation for Born-Infeld electrodynamics has a very
notable form \cite{Chernitskii1998b}. Also, some
propositions which set off Born-Infeld electrodynamics from
electrodynamical models, are presented in \cite{Deser1999}.
About author's opinion on a significance of such nonlinear electrodynamics
 models see \cite{Chernitskii1999}.

At present, Born-Infeld model is considered as a possible
vacuum electrodynamics and future experiments must have solved what
model is more appropriate \cite{Denisov2000}.

Thus, any results which may help us the mathematical investigation of
this field model are strongly welcomed.

For the conventional Maxwell electrodynamics we know the Clifford
number representation (see \cite{Riesz1993,Hestenes1984}).
Maxwell system of equations may be written in highly compact form
for this representation. In addition, it may be more convenient
for some mathematical problems.
Clifford numbers give also the representation of Maxwell system
in massless Dirac equation form, that may help to find more profound
connections between electrodynamics and quantum mechanics.

In this connection the application of Clifford numbers for representation of
Born-Infeld nonlinear electrodynamics model is the important problem.

\section{Clifford numbers and Maxwell equations}
\setcounter{equation}{0}

Clifford numbers are convenient enough
for treatment of geometric objects in space-time.
In particular, we can relate the tensor of electromagnetic field to
the Clifford number by the formula
\be
\hat{F} {}={}
\frac{1}{2}\,F_{\mu\nu}\,\hgamma^\mu\,\hgamma^\nu {}={}
E_i \left(\hgamma^i\wedge\hgamma^0 \right){}+{}
\frac{1}{2}\,\varepsilon_{0ijk}\,B^k\,\hgamma^i\,\hgamma^j
\quad,
\label{Def:hF}
\ee
where the Greek indexes take the values 0,1,2,3, the Latin ones take the
values 1,2,3, and \mbox{$\varepsilon_{\mu\nu\sigma\rho}$} are the
components of fully antisymmetric fourth-rank tensor such that
\mbox{$\varepsilon_{0123} {}={} \sqrt{|g|}$}. Here
we use the designations for electromagnetic field vectors
which are appropriate to arbitrary coordinate system \cite{Chernitskii1998a}.
We designate the Clifford numbers by letters with hat, as distinct from
ordinary numbers.

Symmetric  (inner) and antisymmetric (exterior) Clifford number products
are designated by dot \mbox{($\cdot$)} and
wedge \mbox{($\wedge$)} accordingly.
For the reference vectors \mbox{$\hgamma^\mu$} we have
\mbox{$\hgamma^\mu {}\cdot{} \hgamma^\nu {\,} = {\,}  \hat{1}\,g^{\mu\nu}$}.
They can be represented by Dirac's matrices and
$\hat{1}$ means unit Clifford number or $(4\times4)$ unit matrix.

Notice that \mbox{$\hat{F}$} in (\ref{Def:hF}) is
an invariant geometrical object. This means, in particular, that in
the matrix representation
the appropriate matrix components do not depend on
used coordinate system.

If we introduce the invariant differential operator
\be
\hat{\p} {}\equiv{} \hgamma^\mu\,\dfrac{\p }{\p x^\mu} {}\equiv{}
\hgamma^\mu\,\p_\mu
\quad,
\ee
then the equation
\be
\hp\,\hF  {}={} 0
\label{Eq:MCliff}
\ee
is coordinate-free form of Maxwell system of equations (without currents).
For the specified form of Dirac's matrices we can easily obtain the appropriate
coordinate representation from (\ref{Eq:MCliff}).

We introduce some  designations
concerning Clifford numbers, which we will use:
\be
\ba{lcl}
\hrho^i \;{}\equiv{}\; \hgamma^i {}\wedge{} \hgamma^0
&\quad \Longrightarrow \quad &
 \hrho^i \cdot \hrho^j \;{}={}\;
-g^{00}\,g^{ij} {}+{} g^{i0}\,g^{j0}\quad,\\[7pt]
\hrho^i\cdot\hrho_j \;{}\equiv{}\; \delta^i_j
&\quad\Longrightarrow\quad &  \hrho_i \;{}={}\; \hgamma_0 {}\wedge{} \hgamma_i
\quad,
\ea
\lb{Def:brho}
\ee
where
\mbox{$\hgamma^\mu\cdot\hgamma_\nu \;{}\equiv{}\;\hat{1}\,\delta^\mu_\nu$},
and
\be
\lb{Def:ki}
\ki \equiv  \frac{1}{4!}\,
\varepsilon_{\mu\nu\sigma\rho}\hgamma^\mu\,\hgamma^\nu\,\hgamma^\sigma\,
\hgamma^\rho
\quad
\ee
implies
\be
 \left(\ki \right)^2 {}={} -\hat{1}
\quad,\qquad
 \hgamma^j\,\ki {}={} -\ki\,\hgamma^j
\quad,\qquad
\hrho^j\,\ki {}={} \ki\,\hrho^j
\quad.
\lb{Prop:ki}
\ee

As we see, the Clifford number $\ki$ is coordinate-free form of
fully antisymmetric fourth-rank tensor.  Because for space-time
\mbox{$\ki^2 {}={} -\hat{1}$} we can call it here
Clifford imaginary unit. (This appellation is not allowable for
arbitrary space because there are spaces for which construction of type
(\ref{Def:ki}) has unit square.)

By analogy with ordinary complex numbers, we
introduce the conjugation operation for
Clifford numbers in four-dimensional space-time.
Any such Clifford number or any \mbox{$(4 {}\times{} 4)$} real matrix
 can be represented as
\bse
\bea
\hat{C} &=& a\,\hat{1} {}+{} b_\mu\,\hgamma^\mu {}+{} c_i\,\hrho^i {}+{}
d_i\,\ki\,\hrho^i {}+{} e_\mu\,\ki\,\hgamma^\mu  {}+{} f\,\ki\\
&=& \left(a {}+{} \ki\,f\right)\hat{1} {}+{}
\left(b_\mu {}+{} \ki\,e_\mu\right)\hgamma^\mu {}+{}
\left(c_i {}+{} \ki\,d_\i\right) \hrho^i
\quad,
\lb{Def:calCb}
\eea
\lb{Def:calC}
\ese
where \mbox{$a,\,b_\mu,\,c_i,\,d_i,\,e_\mu,\,f$} are real numbers,
and we define the conjugate Clifford number as
\bse
\bea
\hat{C}^\dagger &=& \left(a {}-{} \ki\,f\right)\hat{1} {}+{}
\left(b_\mu {}-{} \ki\,e_\mu\right)\hgamma^\mu {}+{}
\left(c_i {}-{} \ki\,d_i\right) \hrho^i\\
&=& a\,\hat{1} {}+{} b_\mu\,\hgamma^\mu {}+{} c_i\,\hrho^i {}-{}
d_i\,\ki\,\hrho^i {}-{} e_\mu\,\ki\,\hgamma^\mu  {}-{} f\,\ki
\quad.
\eea
\lb{Def:conjCliff}
\ese
Using (\ref{Def:brho}), (\ref{Def:ki}), (\ref{Prop:ki}), (\ref{Def:calC}), and
(\ref{Def:conjCliff}) we can write
\be
\hF {}={} \hE  {}-{} \ki\,\hB
\quad,\qquad
\hF^\dagger {}={} \hE  {}+{} \ki\,\hB
\quad,
\lb{Def:hFeb}
\ee
where
\be
\lb{Def:EB}
\hE {}\equiv{} E_i\,\hrho^i
\quad,\qquad
\hB {}\equiv{} B^i\,\hrho_i
\quad.
\ee

Clifford number (\ref{Def:calC}) is invariant geometriclal object
and we can mark out three part of it.
In representation  (\ref{Def:calCb}) the first part is scalar because
the coefficients \mbox{$a$} and \mbox{$f$} are scalars for arbitrary
coordinate transformation. The second part is vector and
the coefficients \mbox{$b_\mu$} and \mbox{$e_\mu$} transform like
vector's components. The third part is known as bivector and,
as we see, it is an antisymmetrical second-rank tensor.

In (\ref{Def:calC}) we call \mbox{$a$} as real part of scalar and
\mbox{$\ki\, f$} as imaginary part of it. Also, we mark out
real (\mbox{$b_\mu\,\hgamma^\mu$}) and imaginary
(\mbox{$\ki\,e_\mu\,\hgamma^\mu$}) parts of vector,
and real (\mbox{$c_i\,\hrho^i$}) and imaginary (\mbox{$\ki\, d_i\,\hrho^i$})
parts of bivector.

Equation (\ref{Eq:MCliff}) is written also in the form
\be
\hp {}\cdot{} \hF {}+{} \hp {}\wedge{} \hF {}={} 0
\quad.
\lb{Eq:Maxdw}
\ee
Using representation (\ref{Def:hFeb}), we can write it in the following form:
\be
\hgamma^0\, \p_0 {}\wedge{} \hE {}+{} \hgamma^i\, \p_i {}\cdot{} \hE
{}+{} \hgamma^i\, \p_i {}\wedge{} \hE {}-{}
\hgamma^0\, \p_0 {}\cdot{} \ki\, \hB {}-{}
\hgamma^i\, \p_i {}\cdot{} \ki\, \hB {}-{}
\hgamma^i\, \p_i {}\wedge{} \ki\, \hB {}={} 0
\,.\;
\lb{Eq:Maxwellex}
\ee
Here we use that \mbox{$\hgamma^0 {}\cdot{} \hrho_i {}={} 0$},
\mbox{$\hgamma^0 {}\wedge{} \ki\, \hrho_i {}={} 0$} and
for simplicity we consider that the reference vectors \mbox{$\hgamma^\mu$}
do not depend on coordinates. The left-hand side of (\ref{Eq:Maxwellex})
is a vector. We single out its time and space components,
real and imaginary parts in each of them. As a result, we have the system
\bse
\lb{Eq:MaxwellFull}
\bea
\lb{Eq:MaxwellFulla}
\hgamma^i\, \p_i {}\wedge{} \hE &=& 0
\quad,\\
\lb{Eq:MaxwellFullb}
\hgamma^i\, \p_i {}\cdot{} \ki\, \hB &=& 0
\quad,\\
\lb{Eq:MaxwellFullc}
\hgamma^0\, \p_0 {}\wedge{} \hE {}-{}
\hgamma^i\, \p_i {}\wedge{} \ki\, \hB  &=& 0
\quad,\\
\lb{Eq:MaxwellFulld}
\hgamma^0\, \p_0 {}\cdot{} \ki\, \hB {}-{}
\hgamma^i\, \p_i {}\cdot{} \hE &=& 0
\quad.
\eea
\ese
Here we use (\ref{Def:EB}) and the following relations:
\be
\lb{Tools}
\ba{rclrclrcl}
\hgamma^i {}\wedge{} \hrho_j &=& -\delta^i_j\,\hgamma_0
\;\;,\;\;\;
\hgamma^0 {}\wedge{} \hrho_i &=& \hgamma_i
\;\;,\;\;\;
\hgamma^i {}\wedge{} \ki\, \hrho^j &=& -\varepsilon^{0ijk}\,\hgamma_k
\;\;,\\[1ex]
\hgamma^i {}\cdot{} \ki\, \hrho_j &=& \delta^i_j\,\ki\,\hgamma_0
\;\;,\;\;\;
\hgamma^0 {}\cdot{} \ki\, \hrho_i &=& -\ki\, \hgamma_i
\;\;,\;\;\;
\hgamma^i {}\cdot{} \hrho^j &=& -\varepsilon^{0ijk}\,\ki\,
\hgamma_k
\;\;,
\ea
\ee
where \mbox{$\varepsilon^{0123} {}={} -1/\sqrt{|g|}$}.

We can easily verify that system (\ref{Eq:MaxwellFull}) is Maxwell
system of equations, where
the left-hand sides of (\ref{Eq:MaxwellFulla}) and (\ref{Eq:MaxwellFullb})
are divergences of electromagnetic field vectors.

Now, take a Cartesian coordinate system with
$-g^{00} {}={} g^{11} {}={} g^{22} {}={} g^{33} {}={} 1$ and
$g^{\mu\nu} {}={} 0$ for $\mu\ne\nu$ .
For this case
\mbox{$\ki {}={} \hgamma^0\,\hgamma^1\,\hgamma^2\,\hgamma^3$}.

Consider the appropriate Dirac's matrices in the form:
\be
\ba{rcllrcl}
\hgamma^0 &=&
\left(\matrix{ \i & 0 & 0 & 0 \cr 0 & \i & 0 & 0 \cr
0 & 0 & -\i & 0 \cr 0 & 0 & 0 & -\i     \cr  }\right) &,\;\;&
\hgamma^1 &=&
\left(\matrix{ 0 & 0 & 0 & \i \cr 0 & 0 & \i & 0 \cr
0 & -\i & 0 & 0 \cr -\i & 0 & 0 & 0    \cr  }\right)\;\;, \\[30pt]
\hgamma^2 &=&
\left(\matrix{ 0 & 0 & 0 & -1 \cr 0 & 0 & 1 & 0 \cr
0 & 1 & 0 & 0 \cr -1 & 0 & 0 & 0    \cr  }\right)  &,\;\;&
\hgamma^3 &=&
\left(\matrix{ 0 & 0 & \i & 0 \cr 0 & 0 & 0 & -\i \cr
-\i & 0 & 0 & 0 \cr 0 & \i & 0 & 0    \cr  }\right)\;\;.
\ea
\label{Def:MDir}
\ee

As we can see, the matrix $\hF$ in (\ref{Def:hF}) has four independent
complex components only. We designate the first column of
the matrix $\hF$ by $\bar{F}$, then
\be
\bar{F} {}={} \hF\,\bar{1}
 {} = {}
\left(\matrix{\bar{F}_1\cr
\bar{F}_2\cr \bar{F}_3\cr \bar{F}_4\cr}\right) {} = {}
\left(\matrix{ -\i\,B_3\cr -\i\,B_1 {}-{} B_2\cr E_3\cr E_1
{}-{} \i\,E_2\cr}\right)
\quad,
\qquad
\bar{1} {}\equiv{}
\left(\matrix{1\cr 0\cr 0\cr 0\cr}\right)
\quad.
\label{Def:barF}
\ee
And the matrices $\hF$ and $\hF^\dagger$ can be written in the following form:
\bse
\label{MatrF}
\bea
\label{MatrFa}
\hF &=& \left(\matrix{
\bar{F}_1 &-\bar{F}_2^* &\phantom{-}\bar{F}_3 & \phantom{-}\bar{F}_4^*\cr
\bar{F}_2 &\phantom{-}\bar{F}_1^* &\phantom{-}\bar{F}_4 & -\bar{F}_3^*\cr
\bar{F}_3 &\phantom{-}\bar{F}_4^* &\phantom{-}\bar{F}_1 & -\bar{F}_2^*\cr
\bar{F}_4 & -\bar{F}_3^* &\phantom{-}\bar{F}_2 &\phantom{-}\bar{F}_1^*\cr
}\right)
\quad,
\\
\hF^\dagger &=& \left(\matrix{
-\bar{F}_1 &\phantom{-}\bar{F}_2^* &\phantom{-}\bar{F}_3 & \phantom{-}\bar{F}_4^*\cr
-\bar{F}_2 &-\bar{F}_1^* &\phantom{-}\bar{F}_4 & -\bar{F}_3^*\cr
\phantom{-}\bar{F}_3 &\phantom{-}\bar{F}_4^* &-\bar{F}_1 & \phantom{-}\bar{F}_2^*\cr
\phantom{-}\bar{F}_4 & -\bar{F}_3^* &-\bar{F}_2 &-\bar{F}_1^*\cr
}\right)
\quad.
\label{MatrFb}
\eea
\ese

Extracting the first column in matrix equation (\ref{Eq:MCliff}), we
obtain the following representation of Maxwell equations in
the form of massless Dirac equation:
\be
\hp\,\bar{F} {}={} \hgamma^\mu\,\dfrac{\p\,\bar{F}}{\p x^\mu}  {}={} 0
\quad.
\lb{Eq:MaxwellDirac}
\ee

Note that system (\ref{Eq:MaxwellFull}) is overdetermined
one, because there are eight equations for six unknown functions
\mbox{$E_i(x),B_i(x)$}.
Traditionally,  determined system (\ref{Eq:MaxwellFullc}),
(\ref{Eq:MaxwellFulld})
with additional conditions (\ref{Eq:MaxwellFulla}), (\ref{Eq:MaxwellFullb})
 is considered.
These conditions are conserved in time according to
(\ref{Eq:MaxwellFullc}), (\ref{Eq:MaxwellFulld}). Thus if we
have the problem with initial conditions which satisfy to
(\ref{Eq:MaxwellFulla}), (\ref{Eq:MaxwellFullb}), then
(\ref{Eq:MaxwellFulla}), (\ref{Eq:MaxwellFullb}) will  be
satisfied for any point of time.
But general solution of system
(\ref{Eq:MaxwellFullc}), (\ref{Eq:MaxwellFulld}) is not general solution
of Maxwell equations. Formally, system (\ref{Eq:MaxwellDirac})
has four complex equations for four complex unknown functions
or column \mbox{$\bar{F}(x)$} in (\ref{Def:barF}). But in this case, we have
also two additional conditions that \mbox{$\bar{F}_1$} is pure imaginary
and \mbox{$\bar{F}_3$} is real. Of course, the form of such conditions
is connected with the specified form of matrix representation
for \mbox{$\hgamma^\mu$} which we use.

\section{Coordinate transformations and spinors}
\setcounter{equation}{0}

Because we have Maxwell equations in massless Dirac equation form
(\ref{Eq:MaxwellDirac}), it is necessary to elucidate the
transformation properties of an unknown column \mbox{$\bar{F}$}.
As noted above, because \mbox{$\hgamma^\mu$} are the reference vectors,
the operator \mbox{$\hp$} and the bivector \mbox{$\hF$} are invariant
geometrical objects. Thus if we have two coordinate systems
\mbox{$\{x^\mu\}$} and \mbox{$\{x^{\prime\mu}\}$}, then
\be
\hp
=  \hgamma^{\mu}\,\dfrac{\p}{\p x^{\mu}}
= \hgamma^{'\nu}\,\dfrac{\p}{\p x^{'\nu}}
\quad,\qquad
\hF =  F_{\mu\nu}\,\hgamma^\mu\,\hgamma^\nu  =
F^{\prime}_{\mu\nu}\,\hgamma^{'\mu}\,\hgamma^{'\nu}
\quad.
\lb{hphF}
\ee
Consider the Lorentz transformation
\be
 x'{}^{\nu}  {} = {}  L^\nu_{.\mu}\, x^\mu
\quad.
\lb{Tr:Lor}
\ee
Then for Dirac's matrices we have
\bea
\hgamma^{'\nu}  {} = {}  L^\nu_{.\mu}\,\hgamma^\mu
{} = {} \hLambda^{-1}\,\hgamma^\nu\,\hLambda
\quad,
\lb{Trans}
\eea
where $\hLambda$ is some matrix.

When we relate a specified form of Dirac's matrices (e.g.,
(\ref{Def:MDir})) with appointed coordinate system
\mbox{$\{x^\mu\}$}, then we define invariant
matrix \mbox{$\hF$}, numerical values of its components do not depend
on coordinate system. That is, for example, the component \mbox{$\hF_{11}$}
or \mbox{$\bar{F}_1$} is $-\i\,B_3$.
If we use another coordinate system \mbox{$\{x^{\prime\mu}\}$},
matrices of reference vectors are changed to \mbox{$\{\hgamma^{\prime\mu}\}$}
but it is kept that \mbox{$\hF_{11} {}={} \bar{F}_1 {}={} -\i\,B_3$},
where $B_3$ is third component of magnetic field vector in the
coordinate system \mbox{$\{x^\mu\}$}. Thus column
\mbox{$\bar{F}$} in (\ref{Eq:MaxwellDirac}) is also
invariant object.

In this connection it should recall the article in which
Dirac introduce his equation \cite{Dirac1928}.
In this article he considers the case when unknown column
is invariant. On the other hand,  Pauli \cite{Pauli1936}
has proposed to keep the $\gamma$-matrices invariable
when the coordinates are transformed. In this case, we must
transform the unknown column into Dirac equation. In our
designations this means that
\be
\gamma^\mu\,\frac{\p \bar{F}}{\p x^\mu}  {} = {}  0
\quad\longrightarrow\quad
\gamma^\mu\,\frac{\p \bar{F}^\prime}{\p x^{\prime\mu}}  {} = {}  0
\qquad\Longrightarrow\qquad
\bar{F}^\prime {}={}  \hLambda\,\bar{F}
\quad,
\lb{Tr:Spinor}
\ee
where it is essential that $\hLambda {}\neq{} \hLambda (x)$.

This transformation for column \mbox{$\bar{F}$} is called
spinor transformation. On this topic see
\cite{Sommerfeld1951}, where comparative analysis for these two approaches
to the invariance of Dirac equation in respect of Lorentz transformations
is given.

Though in (\ref{Tr:Spinor}) we have Lorentz transformation for
coordinates (\ref{Tr:Lor}), transformation (\ref{Tr:Spinor})
is not space-time rotation for Maxwell equations in massless Dirac equation
form (\ref{Eq:MaxwellDirac}).
Really, so far as we keep relation between components of \mbox{$\bar{F}$}
and $E_i,B_i$ (\ref{Def:barF}),  transformation (\ref{Tr:Spinor}) changes
the vector components $E_i,B_i$ to components of
some another vector $E_i^\prime,B_i^\prime$ which, in general, may be
complex numbers.
But transformation (\ref{Tr:Spinor}) keeps (\ref{Eq:MaxwellDirac})
to be invariant. Hence this transformation give new solution
of (\ref{Eq:MaxwellDirac}):
\be
\bar{F} ( x^\mu ) {\,}\longrightarrow {\,}
\hLambda\,\bar{F} ( L^\mu_{.\nu} x^\nu )
\quad.
\lb{Sol:TransSpin}
\ee
Here the matrices $\hLambda$ and $(L^\mu_{.\nu})$ are interconsistent
in the sense of the relation
\mbox{$L^\nu_{.\mu}\,\hgamma^\mu {} = {}
\hLambda^{-1}\,\hgamma^\nu\,\hLambda$},
but transformation (\ref{Sol:TransSpin}) is not real space-time rotation.
In particular, if Lorentz matrix $(L^\mu_{.\nu})$ corresponds to
the space rotation for \mbox{$2\,\pi$}, then transformation (\ref{Sol:TransSpin})
change any solution of Maxwell equations to the same solution
with opposite sign.

Equation (\ref{Eq:MaxwellDirac}) may be called spinor
representation for Maxwell equations.

\section{Born-Infeld electrodynamics and\\ Clifford numbers}
\setcounter{equation}{0}

Equations of Born-Infeld electrodynamics in arbitrary coordinates
(and outside of singularities) are written in the following
form (see \cite{Chernitskii1998a}):
\bse
\lb{Eq:NE}
\bea
\sqr^{-1}\,\p_i\,\sqr\,D^i &=& 0
\quad,
\\
\sqr^{-1}\,\p_i\,\sqr\,B^i &=& 0
\quad,
\\
\sqr^{-1}\,\p_0\,\sqr\,D^i {}+{} \varepsilon^{0ijk}\,\p_j\,H_k &=& 0
\quad,
\\
\sqr^{-1}\,\p_0\,\sqr\,B^i {}-{} \varepsilon^{0ijk}\,\p_j\,E_k &=& 0
\quad,
\eea
\ese
where  $\;E_i {}\equiv{} F_{i0}$ ,
$\;B^i {}\equiv{}  -\dfrac{1}{2}\, \varepsilon^{0ijk}\, F_{jk}$ ,
$\;D^i {}\equiv{}  f^{0i}$ ,
$\;H_i {}\equiv{} \dfrac{1}{2}\,\varepsilon_{0ijk}\, f^{jk}$ ,\\ and
\bea
&&f^{\mu\nu}  {}={}
\dfrac{1}{\L}\left[F^{\mu\nu}  {}-{}
\dfrac{\chi^2}{2}\,\J\,\varepsilon^{\mu\nu\sigma\rho}\,F_{\sigma\rho}\right]
\quad,
\label{Def:f}
\\
& & \L {}\equiv{}
\sqrt{|\,1 {}-{}  \chi^2\,\I  {}-{}  \chi^4\,\J^2\,|}\;,\;\;
\I {}\equiv{}  \dfrac{1}{2}\,F_{\mu\nu}\,F^{\nu\mu}
\;,\;\;
\J {}\equiv{}
\dfrac{1}{8}\,\varepsilon_{\mu\nu\sigma\rho}\, F^{\mu\nu} F^{\sigma\rho}
\;,
\non
\eea
\mbox{$\chi$} is some dimensional constant.

By analogy with (\ref{Def:hF}) and (\ref{Def:hFeb}) we have
\be
\hat{f} {}={}
\frac{1}{2}\,f^{\mu\nu}\,\hgamma_\mu\,\hgamma_\nu {}={}
\hat{D} {}-{} \ki\,\hat{H}
\quad.
\label{Def:hf}
\ee

Now consider the following equation:
\be
\hp {}\cdot{} \hF {}+{} \hp {}\wedge{} \hat{f} {}={} 0
\lb{Eq:NEdw}
\quad.
\ee
By analogy with derivation from (\ref{Eq:Maxdw}) to (\ref{Eq:MaxwellFull}),
we have from (\ref{Eq:NEdw}) that
\bse
\lb{Eq:NEFull}
\bea
\hgamma^i\, \p_i {}\wedge{} \hat{D} &=& 0
\quad,\\
\hgamma^i\, \p_i {}\cdot{} \ki\, \hB &=& 0
\quad,\\
\hgamma^0\, \p_0 {}\wedge{} \hat{D} {}-{}
\hgamma^i\, \p_i {}\wedge{} \ki\, \hat{H}  &=& 0
\quad,\\
\hgamma^0\, \p_0 {}\cdot{} \ki\, \hB {}-{}
\hgamma^i\, \p_i {}\cdot{} \hE &=& 0
\quad.
\eea
\ese
This system conforms with system (\ref{Eq:NE}) in the coordinates
for which \mbox{$\hgamma^\mu {}\neq{} \hgamma^\mu (x)$}.
But coordinate-free equation (\ref{Eq:NEdw}) fully conforms
with system (\ref{Eq:NE}) for any coordinates. This is also verified
by direct substitution.

As we can see, system (\ref{Eq:NE}) is appropriate
for any nonlinear electrodynamics model. Thus, (\ref{Eq:NEdw})
is a Clifford number form for equations of nonlinear electrodynamics.
In this case the relation between $\hat{f}$ and $\hF$ defines concrete model.
It can be easily shown that Born-Infeld relation (\ref{Def:f}) is written
in the form
\bea
\lb{Def:hffromhF}
&&\hat{f} {}={} \dfrac{1}{\L}\left(\hat{1} {}+{} \ki\,\chi^2\,\J\right)\hF
\quad,
\\
&&
\I\,\hat{1} {}={}  \dfrac{1}{2}\left[\hF^2 {}+{} (\hF^\dagger)^2\right]
\;,\;\;
\J\,\hat{1} {}={}
\dfrac{1}{4}\,\ki\left[\hF^2  {}-{} (\hF^\dagger)^2\right]
\;.
\non
\eea

Making elementary transformations with (\ref{Eq:NEdw}) and
extracting the first column in matrix equation (see (\ref{Def:barF}) and
(\ref{Eq:MaxwellDirac})), we obtain
\be
\hp\,\bar{F} {}={}
\left[ \hp {}\wedge{} \left(\hF {}-{} \hat{f}\right)\right] \bar{1}
\quad.
\lb{Eq:NEDirac}
\ee
Using relation of type (\ref{Def:hffromhF}) and representation
(\ref{MatrF}) the right-hand side of
(\ref{Eq:NEDirac}) can be represented by \mbox{$\bar{F}$} and
\mbox{$\bar{F}^{*}$}. Thus, we have the nonlinear like Dirac equation
which is the form of nonlinear electrodynamics equations.

\section{Conclusion}
\setcounter{equation}{0}

We have considered the Clifford number formalism for Maxwell
equations.
The Clifford imaginary unit for space-time \mbox{$\ki$} is introduced
as coordinate
independent form of fully antisymmetric fourth-rank tensor. Earlier
(see \cite{Hestenes1984}) the pseudoscalar
\mbox{$\hgamma^0\,\hgamma^1\,\hgamma^2\,\hgamma^3$} was used as
such imaginary unit. This distinction plays an important role for
curvilinear coordinates and for generalization to curved space-time.

We have considered the representation of Maxwell equations in massless
Dirac equation form. In this connection, we have discussed two approaches
to the invariance of Dirac equation in respect of the Lorentz transformations.
According to the
first unjustly forgotten approach, the unknown column is invariant
and according to the second
approach, it has the transformation properties known as spinorial ones.

We have obtained the Clifford number representation for nonlinear
electrodynamics equations. From this representation, we obtain
the nonlinear like Dirac equation which is the form of nonlinear
electrodynamics equations.
As the special case, we have obtained the appropriate representations
for Born-Infeld nonlinear electrodynamics.

These representations can help
to obtain new solutions in Born-Infeld electrodynamics.
Also such approach may make clear the relations between nonlinear
electrodynamics and particle physics.

\end{document}